\documentclass[preprint,12pt]{elsarticle}
\usepackage{amssymb}
\usepackage{amsmath}

\journal{Parallel Computing}

\usepackage{enumerate}
\usepackage{xspace}
\usepackage[noend]{algpseudocode}
\usepackage{algorithm}
\usepackage{color}

\newcommand{\PRAGMATIC}{PRAgMaTIc\xspace}
\newcommand{\ie}{{\em i.e.}\xspace}
\newcommand{\eg}{{\em e.g.}\xspace}
\newcommand{\etal}{{\em et al.}\xspace}

\let\And\undefined
\newcommand{\And}{\textbf{and}}
\newcommand{\Or}{\textbf{or}}

\newcommand{\Figref}[1]{Figure~\ref{#1}}
\newcommand{\fig}[4]{
 \begin{figure}[t]
 \begin{center}
 \includegraphics[width=#3\linewidth]{#1}
 \caption[ ]{\label{#4} #2}
 \end{center}
 \end{figure}
}

\begin{document}

\begin{frontmatter}

\title{A thread-parallel algorithm for anisotropic mesh adaptation}

\author[CS]{G.~Rokos}
\author[ESE]{G.J.~Gorman\corref{cor1}}
\ead{g.gorman@imperial.ac.uk}
\author[FLE]{J.~Southern}
\author[ESE]{P.H.J.~Kelly}

\address[CS]{Department of Computing, Imperial College
London, SW7 2AZ, UK}
\address[ESE]{Applied Modelling and Computation Group,
  Department of Earth Science and Engineering, Imperial College
London, SW7 2AZ, UK}
\address[FLE]{Fujitsu Laboratories of Europe, Hayes Park Central,
Hayes End Road, Hayes, Middlesex, UB4 8FE, UK}

\cortext[cor1]{Corresponding author}

\begin{abstract}

Anisotropic mesh adaptation is a powerful way to directly minimise the
computational cost of mesh based simulation. It is particularly
important for multi-scale problems where the required number of
floating-point operations can be reduced by orders of magnitude
relative to more traditional static mesh approaches.

Increasingly, finite element and finite volume codes are
being optimised for modern multi-core architectures.
Typically, decomposition methods implemented through the
Message Passing Interface (MPI) are applied for inter-node
parallelisation, while a threaded programming model, such as OpenMP, is used
for intra-node parallelisation. Inter-node parallelism for mesh
adaptivity has been successfully implemented by a number of
groups. However, thread-level parallelism is significantly more challenging
because the underlying data structures are extensively
modified during mesh adaptation and a greater degree of
parallelism must be realised.

In this paper we describe a new thread-parallel algorithm for
anisotropic mesh adaptation algorithms. For each of the mesh
optimisation phases (refinement, coarsening, swapping and smoothing)
we describe how independent sets of tasks are defined. We show how a
deferred updates strategy can be used to update the mesh data
structures in parallel and without data contention. We show that
despite the complex nature of mesh adaptation and inherent load
imbalances in the mesh adaptivity, a parallel efficiency of 60\% is
achieved on an 8 core Intel Xeon Sandybridge, and a 40\% parallel
efficiency is achieved using 16 cores in a 2 socket Intel Xeon
Sandybridge ccNUMA system.

\end{abstract}

\begin{keyword}
anisotropic mesh adaptation, finite element analysis,
multi-core, OpenMP
\end{keyword}

\end{frontmatter}

\section{Introduction}

Resolution in time and space are often the limiting factors in
achieving accurate simulations for real world problems across a wide
range of applications in science and engineering. The brute force
strategy is typical, whereby resolution is increased throughout the
domain until the required accuracy is achieved. This is successful up
to a point when a numerical method is relatively straightforward to
scale up on parallel computers, for example finite difference or
lattice Boltzmann methods. However, this also leads to a large
increase in the computational requirements. In practice, this means
simulation accuracy is usually determined by the available
computational resources and an acceptable time to solution rather than
the actual needs of the problem.

Finite element and finite volume methods on unstructured
meshes offer significant advantages for many real world
applications. For example, meshes comprised of simplexes
that conform to complex geometrical boundaries can now be
generated with relative ease. In addition, simplexes are
well suited to varying the resolution of the mesh throughout
the domain, allowing for local coarsening and refinement of
the mesh without hanging nodes. It is common for these codes
to be memory bound because of the indirect addressing and the
subsequent irregular memory access patterns that
the unstructured data structures introduce. However,
discontinuous Galerkin and high order finite element methods
are becoming increasingly popular because of their numerical
properties and associated compact data structure
allows data to be easily streamed on multi-core
architectures.

A difficulty with mesh based modelling is that the mesh is generated
before the solution is known, however, the local error in the solution
is related to the local mesh resolution.  The practical consequence of
this is that a user may have to vary the resolution at the mesh
generation phase and rerun simulation several times before a
satisfactory result is achieved. However, this approach is
inefficient, often lacks rigour and may be completely impractical for
multiscale time-dependent problems where superfluous computation may
well be the dominant cost of the simulation.

Anisotropic mesh adaptation methods provide an important means to
minimise superfluous computation associated with over resolving the
solution while still achieving the required accuracy, \eg
\cite{buscaglia1997anisotropic, agouzal1999adaptive,
  pain_tetrahedral_2001, li20053d}. In order to use mesh adaptation
within a simulation, the application code requires a method to
estimate the local solution error. Given an error estimate it is then
possible to calculate the required local mesh resolution in order to
achieve a specific solution accuracy.

There are a number of examples where this class of adaptive
mesh methods has been extended to distributed memory
parallel computers. The main challenge in performing mesh
adaptation in parallel is maintaining a consistent mesh
across domain boundaries. One approach is to first lock the
regions of the mesh which are shared between processes and
for each process to apply the serial mesh adaptation
operation to the rest of the local domain. The domain
boundaries are then perturbed away from the locked region
and the lock-adapt iteration is repeated
\cite{coupez2000parallel}. Freitag \etal
\cite{freitag1995cient, Freitag98thescalability} considered
a fine grained approach whereby a global task graph is defined
which captures the data dependencies for a particular mesh
adaptation kernel. This graph is coloured in order to
identify independent sets of operations. The parallel
algorithm then progresses by selecting an independent set
(vertices of the same colour) and applying mesh adaptation
kernels to each element of the set. Once a sweep through a
set has been completed, data is synchronised between
processes, and a new independent set is selected for
processing. In the approach described by Alauzet \etal
\cite{alauzet2006parallel}, each process applies the serial
adaptive algorithm, however rather than locking the halo
region, operations to be performed on the halo are first
stashed in buffers and then communicated so that the same
operations will be performed by all processes that share
mesh information. For example, when coarsening is applied
all the vertices to be removed are computed. All operations
which are local are then performed while pending operations
in the shared region are exchanged. Finally, the pending
operations in the shared region are applied.

However, over the past decade there has been a trend towards
multi-core compute nodes.  Indeed, it is assumed that the
compute nodes of a future exascale supercomputer will each
contain thousands or even tens of thousands of cores
\cite{dongarra2011exascale}. On multi-core architectures, a
popular parallel programming paradigm is to use thread-based
parallelism, such as OpenMP, to exploit shared memory within
a shared memory node and a message passing using MPI for
interprocess communication. When the computational intensity
is sufficiently high, a third level of parallelisation may
be implemented via SIMD instructions at the core level.
There are some opportunities to improve performance and
scalability by reducing communication needs, memory
consumption, resource sharing as well as improved load
balancing \cite{rabenseifner2009hybrid}.  However, the
algorithms themselves must also have a high degree of thread
parallelism if they are to have a future on multi-core
architectures; whether it be CPU or coprocessor based.

In this paper we take a fresh look at the anisotropic adaptive mesh
methods in 2D to develop new scalable thread-parallel algorithms
suitable for modern multi-core architectures. We show that despite the
irregular data access patterns, irregular workload and need to rewrite
the mesh data structures; good parallel efficiency can be
achieved. The key contributions are:
\begin{itemize}
\item The first threaded implementation of anisotropic mesh adaptation.
\item Demonstration of parallel efficiencies of 60\% on 8 core UMA
  architecture and 40\% on 16 core ccNUMA architecture.
\item Algorithms for selection of maximal independent sets for each
  sweep of mesh optimisation.
\item Use of deferred update to avoid data contention and to
  parallelise updates to mesh data structures.
\end{itemize}

The algorithms described in this paper have been implemented in the
open source code \PRAGMATIC (Parallel anisotRopic Adaptive Mesh
ToolkIt)\footnote{https://launchpad.net/pragmatic}. PRAgMaTIc also
includes MPI parallelisation, which, for space reasons, will be
covered in a forthcoming paper. The remainder of the paper is laid
out as follows: \S \ref{sect:overview} gives an overview of the
anisotropic adaptive mesh procedure; \S \ref{sect:parallel} describes
the thread-parallel algorithm; and \S \ref{sect:benchmark} illustrates
how well the algorithm scales for a benchmark problem. We conclude
with a discussion on future work and possible implications of this
work.

\section{Overview}
In this section we will give an overview of anisotropic mesh
adaptation. In particular, we focus on the element quality as defined
by an error metric and the anisotropic adaptation kernels which
iteratively improve the local mesh quality as measured by the worst
local element.

\label{sect:overview}
\subsection{Error control}
\label{subsect:error_control}

Solution discretisation errors are closely related to the size and the
shape of the elements. However, in general meshes are generated using
{\em a priori} information about the problem under consideration when
the solution error estimates are not yet available. This may be
problematic because:
\begin{itemize}
\item Solution errors may be unacceptably high.
\item Parts of the solution may be over-resolved, thereby incurring
  unnecessary computational expense.
\end{itemize}
A solution to this is to compute appropriate local error estimates,
and use this information to compute a field on the mesh which
specifies the local mesh resolution requirement. In the most general
case this is a metric tensor field so that the resolution requirements
can be specified anisotropically; for a review of the procedure see
\cite{frey2005anisotropic}. Size gradation control can be applied to
this metric tensor field to ensure that there are not abrupt changes
in element size \cite{alauzet2010size}.

\subsection{Element quality}
\label{subsect:element_quality}

As discretisation errors are dependent upon element shape as well as
size, a number of different measures of element quality have been
proposed, \eg \cite{buscaglia1997anisotropic, vasilevskii1999adaptive,
  agouzal1999adaptive, tam2000anisotropic, pain_tetrahedral_2001}.
For the work described here, we use the element quality measure for
triangles proposed by Vasilevskii \etal
\cite{vasilevskii1999adaptive}:
\begin{equation}\label{eqn:quality2D}
q^M(\triangle) = \underbrace{12\sqrt{3}\frac{|\triangle|_M}{|\partial\triangle|_M^2}}_{I} \underbrace{F\left(\frac{|\partial\triangle|_M}{3}\right)}_{II},
\end{equation}
where $|\triangle|_M$ is the area of element $\triangle$ and
$|\partial\triangle|_M$ is its perimeter as measured with respect to
the metric tensor $M$ as evaluated at the element's centre. The first
factor ($I$) is used to control the shape of element $\triangle$. For
an equilateral triangle with sides of length $l$ in metric space,
$|\triangle| = l^2\sqrt{3}/4$ and $|\partial\triangle| = 3l$; and so
$I=1$. For non-equilateral triangles, $I<1$. The second factor ($II$)
controls the size of element $\triangle$. The function $F$ is smooth
and defined as:
\begin{equation}
F(x) = \left(\min(x, 1/x)(2 - \min(x, 1/x))\right)^3,
\end{equation}
which has a single maximum of unity with $x=1$ and decreases smoothly
away from this with $F(0)=F(\infty)=0$. Therefore, $II=1$ when the sum
of the lengths of the edges of $\triangle$ is equal to $3$, \eg\ an
equilateral triangle with sides of unit length, and $II<1$
otherwise. Hence, taken together, the two factors in
(\ref{eqn:quality2D}) yield a maximum value of unity for an
equilateral triangle with edges of unit length, and decreases smoothly
to zero as the element becomes less ideal.

\subsection{Overall adaptation procedure}
\label{subsect:overall_procedure}

The mesh is adapted through a series of local operations: edge
collapse (\S\ref{subsect:coarsening}); edge refinement
(\S\ref{subsect:refinement}); element-edge swaps
(\S\ref{subsect:swapping}); and vertex smoothing
(\S\ref{subsect:smoothing}). While the first two of these operations
control the local resolution, the latter two operations are used to
improve the element quality.

Algorithm \ref{alg:general} gives a high level view of the anisotropic
mesh adaptation procedure as described by \cite{li20053d}.  The inputs
are $\mathcal{M}$, the piecewise linear mesh from the modelling
software, and $\mathcal{S}$, the node-wise metric tensor field which
specifies anisotropically the local mesh resolution
requirements. Coarsening is initially applied to reduce the subsequent
computational and communication overheads. The second stage involves
the iterative application of refinement, coarsening and mesh swapping
to optimise the resolution and the quality of the mesh. The algorithm
terminates once the mesh optimisation algorithm converges or after a
maximum number of iterations has been reached. Finally, mesh quality
is fine-tuned using some vertex smoothing algorithm (e.g.
quality-constrained Laplacian smoothing \cite{freitag1996comparisonw},
optimisation-based smoothing \cite{freitag1995cient}), which aims
primarily at improving worst-element quality.

\begin{algorithm}[t]
  \caption{Mesh optimisation procedure.}
  \label{alg:general}
  \begin{algorithmic}
	\State Inputs: $\mathcal{M}$, $\mathcal{S}$.
	\State $(\mathcal{M}^*, \mathcal{S}^*) \gets coarsen(\mathcal{M}$, $\mathcal{S})$
	\Repeat
	  \State $(\mathcal{M}^*, \mathcal{S}^*) \gets refine(\mathcal{M}^*$, $\mathcal{S}^*)$
	  \State $(\mathcal{M}^*, \mathcal{S}^*) \gets coarsen(\mathcal{M}^*$, $\mathcal{S}^*)$
	  \State $(\mathcal{M}^*, \mathcal{S}^*) \gets swap(\mathcal{M}^*$, $\mathcal{S}^*)$
	\Until{(maximum number of iterations reached) \Or (algorithm convergence)}
    \State $(\mathcal{M}^*, \mathcal{S}^*) \gets smooth(\mathcal{M}^*$, $\mathcal{S}^*)$
    \State \Return $\mathcal{M}^*$
  \end{algorithmic}
\end{algorithm}

\subsection{Adaptation kernels}

\subsubsection{Coarsening}
\label{subsect:coarsening}

Coarsening is the process of lowering mesh resolution locally by
removing mesh elements, leading to a reduction in the computational
cost. Here this is done by collapsing an edge to a single vertex,
thereby removing all elements that contain this edge. An example of
this operation is shown in \Figref{fig:edge_collapse}.

\fig{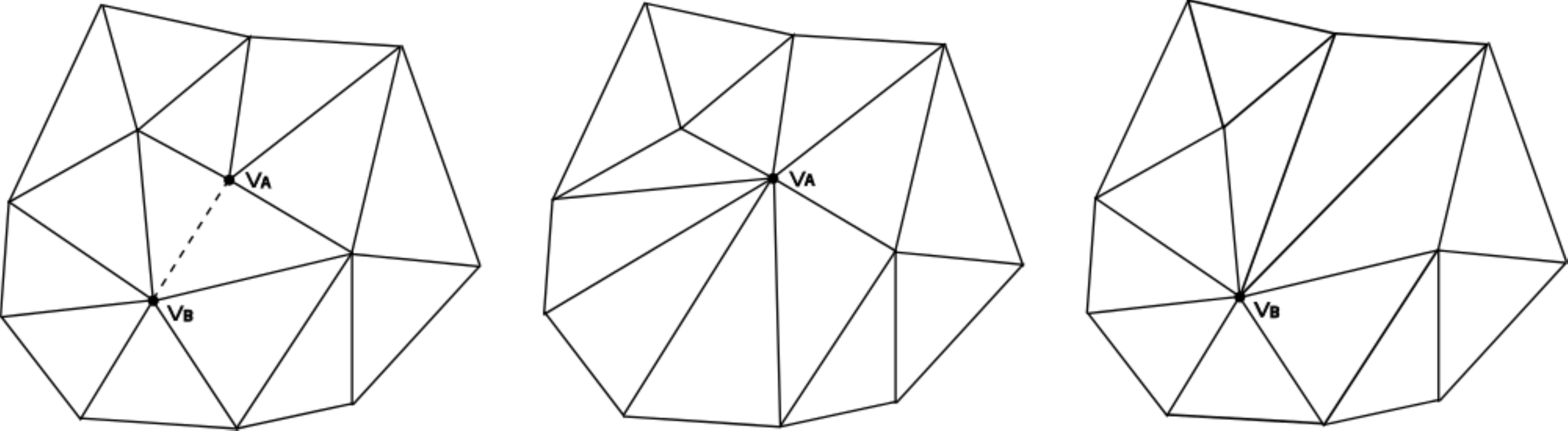}{Edge collapse: The dashed edge in
  the left figure is reduced into a single vertex by bringing vertex
  $V_B$ on top of vertex $V_A$, as can be seen in the middle
  figure. The two elements that used to share the dashed edge are
  deleted. Edge collapse is an oriented operation, \ie eliminating
  the edge by moving $V_B$ onto $V_A$ results in a different local
  patch than moving $V_A$ onto $V_B$, which can be seen in the right
  figure.}{1.0}{fig:edge_collapse}

\subsubsection{Refinement}
\label{subsect:refinement}

Refinement is the process of increasing mesh resolution locally. It
encompasses two operations: splitting of edges; and subsequent
division of elements. When an edge is longer than desired, it is
bisected. An element can be split in three different ways, depending
on how many of its edges are bisected:
\begin{enumerate}
\item When only one edge is marked for refinement, the element can be
  split across the line connecting the mid-point of the marked edge
  and the opposite vertex. This operation is called bisection and an
  example of it can be seen on the left side of Figure
  \ref{fig:refinement} (left shape).
\item When two edges are marked for refinement, the element is divided
  into three new elements. This case is shown in \Figref{fig:refinement}
  (middle shape). The parent element is split by
  creating a new edge connecting the mid-points of the two marked
  edges. This leads to a newly created triangle and a non-conforming
  quadrilateral. The quadrilateral can be split into two conforming
  triangles by dividing it across one of its diagonals, whichever is
  shorter.
\item When all three edges are marked for refinement, the element is
  divided into four new elements by connecting the mid-points of its
  edges with each other. This operation is called regular refinement
  and an example of it can be seen in Figure \ref{fig:refinement}
  (right shape).
\end{enumerate}

\fig{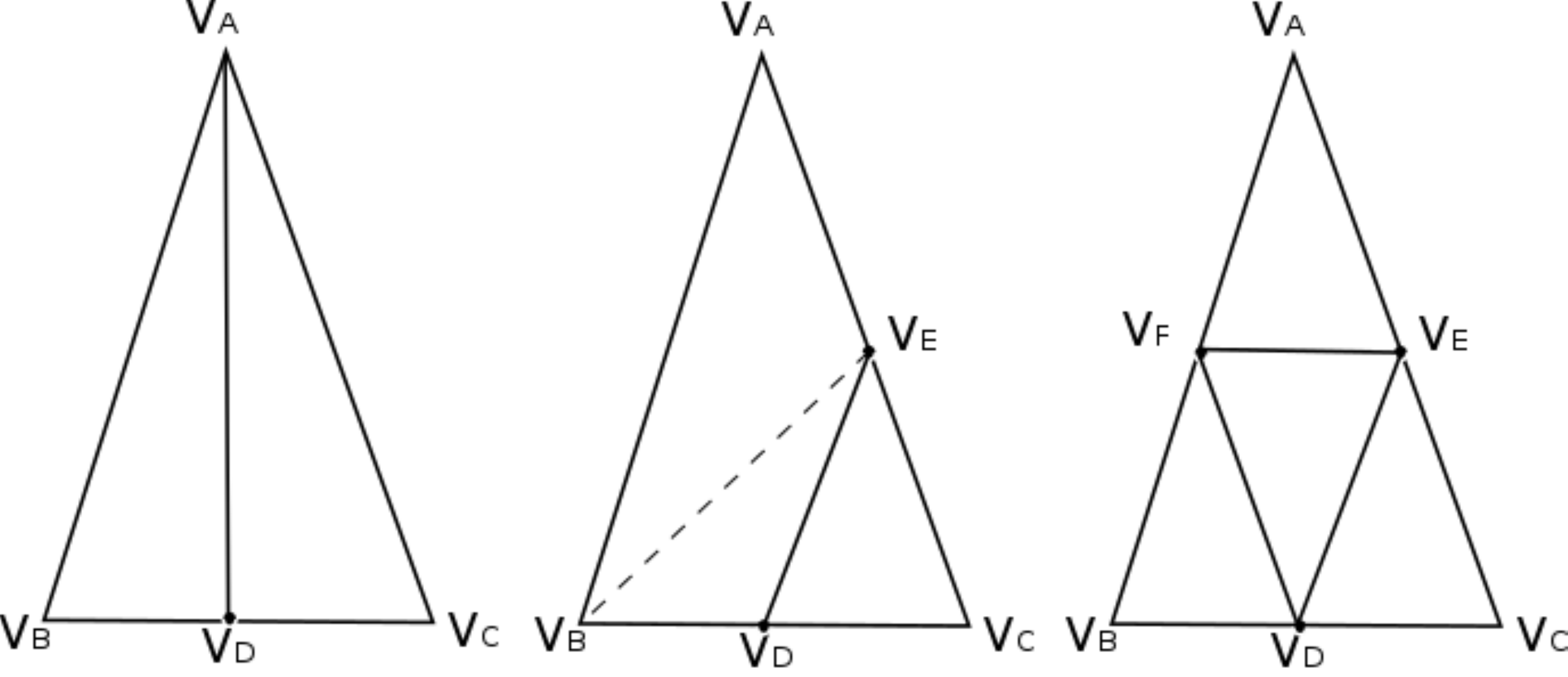}{Mesh resolution can be increased either by
  bisecting an element across one of its edges (1:2 split, left
  figure), by performing a 1:3 split (middle figure) or by
  performing regular refinement to that element (1:4 split, right
  figure).}{0.8}{fig:refinement}

\subsubsection{Swapping}
\label{subsect:swapping}

In 2D, swapping is done in the form of edge flipping, \ie flipping an
edge shared by two elements, \eg \Figref{fig:swapping}. The operation
considers the quality of the swapped elements - if the minimum element
quality has improved then the original mesh triangles are replaced
with the edge swapped elements.

Whereas in refinement, propagation is necessary in order to eliminate
nonconformities, swapping operations may also propagate because
topological changes might give rise to new configurations of better
quality. An illustration of this is shown in
\Figref{fig:swapping_propagation}. After an edge has been flipped,
the local topology is modified and adjacent edges which were not
considered for flipping before are now candidates. This procedure
results in a Delaunay triangulation, \ie one in which the minimum
element angle in the mesh is the largest possible one with respect to
all other triangulations of the mesh \cite{KulkarniBCP09}.

\fig{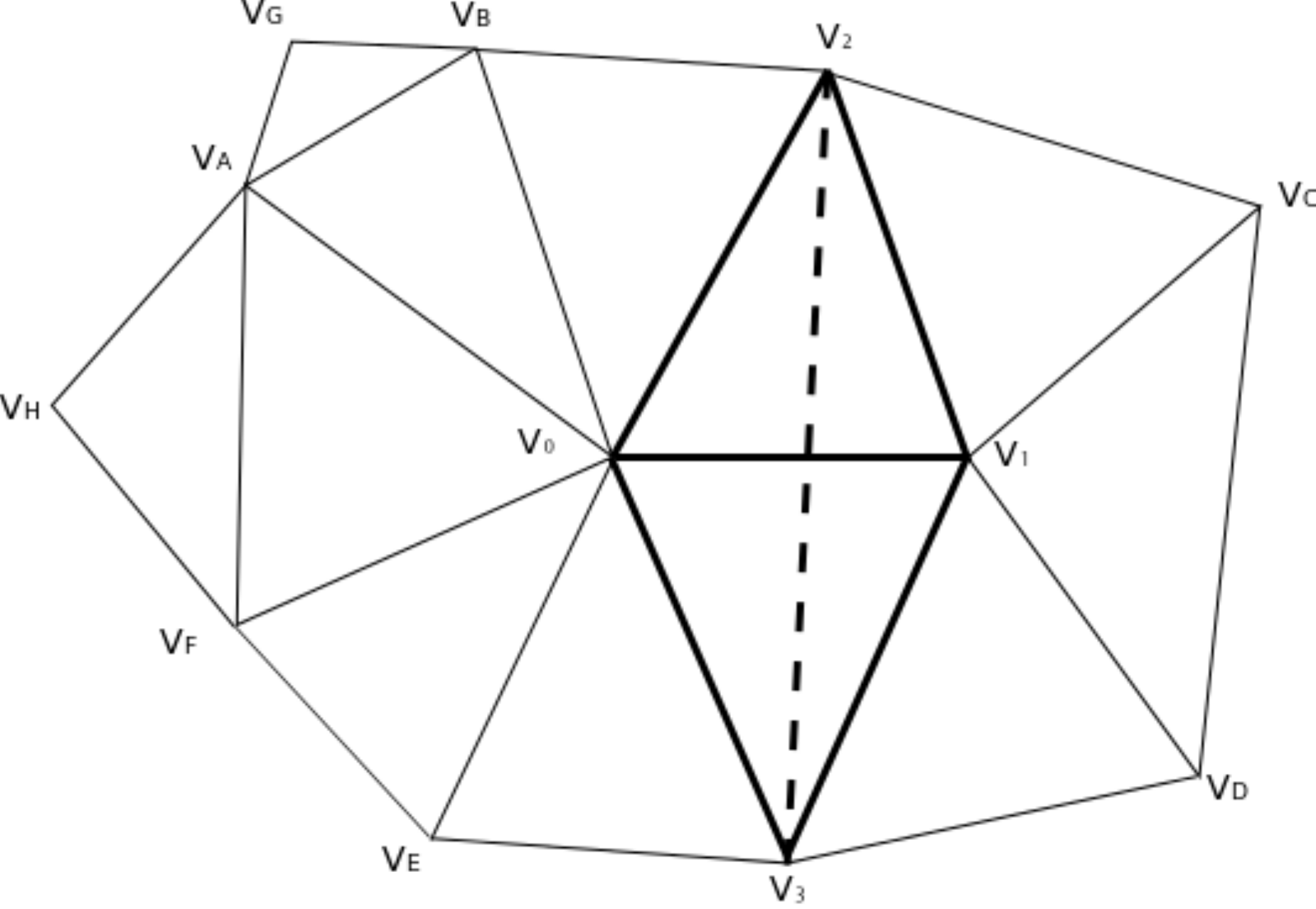}{Flipping the common edge $\overline{V_0 V_1}$
  results in the removal of triangles $\widehat{V_0 V_1 V_2}$ and
  $\widehat{V_0 V_1 V_3}$ and their replacement with new triangles
  $\widehat{V_0 V_2 V_3}$ and $\widehat{V_1 V_2
    V_3}$.}{0.6}{fig:swapping}

\fig{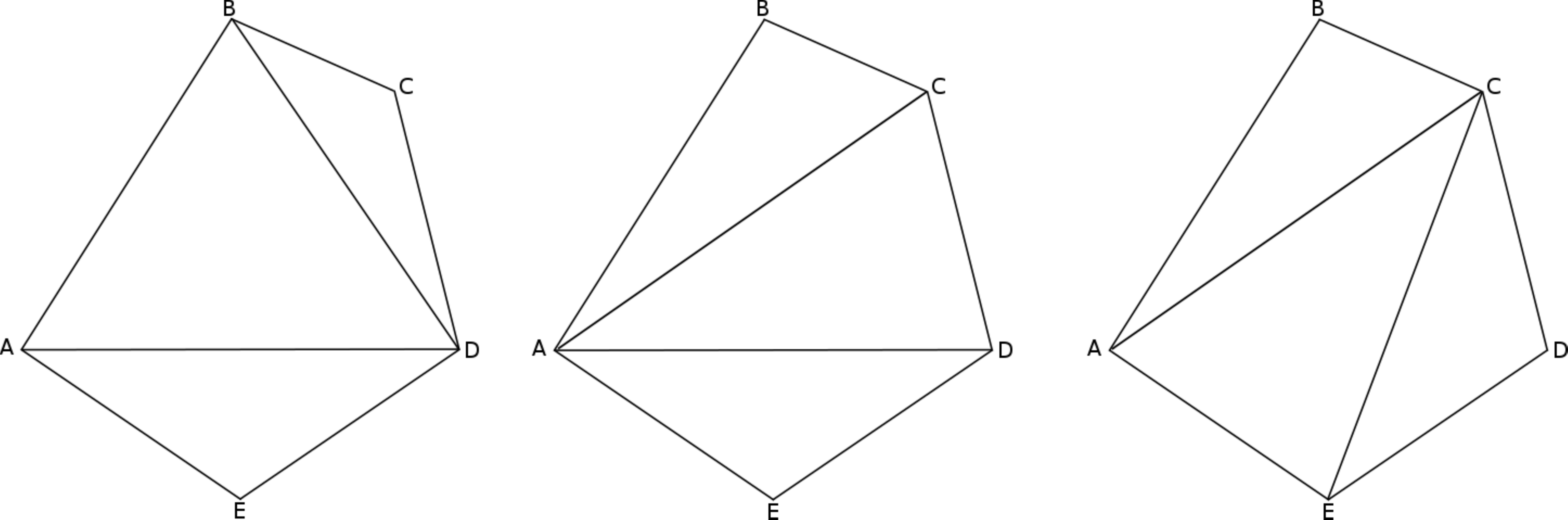}{Initially (left figure), edge $\overline{AD}$
  is not considered a candidate for swapping because the hypothetical
  triangles $\widehat{ABE}$ and $\widehat{BDE}$ are of poorer quality
  than the original triangles $\widehat{ABD}$ and
  $\widehat{ADE}$. Edge $\overline{BD}$, on the other hand, can be
  flipped, resulting in improved quality of the local patch (middle
  figure). After this step, edge $\overline{AD}$ becomes a candidate
  for swapping, as new elements $\widehat{ACE}$ and $\widehat{CDE}$
  are indeed of higher quality than the original elements
  $\widehat{ABD}$ and $\widehat{ADE}$(right
  figure).}{1.0}{fig:swapping_propagation}

\subsubsection{Quality constrained Laplacian Smooth}
\label{subsect:smoothing}

\begin{figure}[t]
\begin{center}
\includegraphics[width=0.5\textwidth]{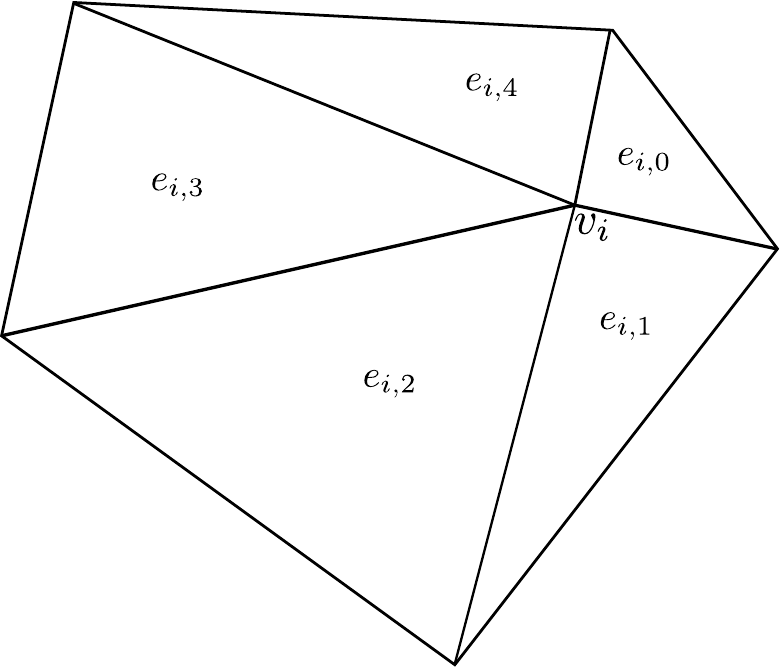}
\end{center}
\caption{Local mesh patch: $\vec{v}_i$ is the vertex being
  relocated; $\{e_{i,0},\ldots,e_{i, m}\}$ is the set of elements
  connected to $\vec{v}_i$.}
\label{fig:patch}
\end{figure}

The kernel of the vertex smoothing algorithm should relocate the
central vertex such that the local mesh quality is increased (see
Figure \ref{fig:patch}). Probably the best known heuristic for mesh
smoothing is Laplacian smoothing, first proposed by Field
\cite{field_laplacian_1988}. This method operates by moving a vertex
to the barycentre of the set of vertices connected by a mesh edge to
the vertex being repositioned. The same approach can be implemented
for non-Euclidean spaces; in that case all measurements of length and
angle are performed with respect to a metric tensor field that
describes the desired size and orientation of mesh elements
(\eg\ \cite{pain_tetrahedral_2001}). Therefore, in general, the
proposed new position of a vertex $\vec{v}_i^{\mathcal{L}}$ is given
by
\begin{equation}\label{eqn:smooth}
  \vec{v}_i^{\mathcal{L}} = \frac{\sum_{j=1}^J \vert\vert {\vec{v}_i - \vec{v}_j}
  \vert\vert_M \vec{v}_j }{\sum_{j=1}^J \vert\vert {\vec{v}_i - \vec{v}_j} \vert\vert_M },
\end{equation}
where $\vec{v}_j$, $j=1,\ldots ,J$, are the vertices connected to
$\vec{v}_i$ by an edge of the mesh, and $\vert\vert\cdot\vert\vert_M$ is
the norm defined by the edge-centred metric tensor
$M_{ij}$. In Euclidean space, $M_{ij}$ is
the identity matrix.

As noted by Field \cite{field_laplacian_1988}, the application of pure
Laplacian smoothing can actually decrease useful local element quality
metrics; at times, elements can even become inverted. Therefore,
repositioning is generally constrained in some way to prevent local
decreases in mesh quality. One variant of this, termed \emph{smart
Laplacian smoothing} by Freitag \etal\ \cite{freitag1996comparisonw},
is summarised in
Algorithm \ref{alg:smart_smooth_kernel} (while Freitag \etal\ only 
discuss this for Euclidean geometry it is straightforward to extend to 
the more general case). This method accepts the new
position defined by a Laplacian smooth only if it increases the
infinity norm of local element quality, $Q_i$ (\ie\ the quality of the
worst local element):
\begin{equation}\label{eqn:qinf}
  Q(\vec{v}_i) \equiv \| \boldsymbol{q} \|_{\infty}
\end{equation}
where $i$ is the index of the vertex under consideration and
$\boldsymbol{q}$ is the vector of the element qualities from the local
patch.

\begin{algorithm}[t]
  \caption{Smart smoothing kernel: a Laplacian smooth operation is accepted
           only if it does not reduce the infinity norm of local element quality.}
  \label{alg:smart_smooth_kernel}
  \begin{algorithmic}
    \State $\vec{v}_i^0 \gets \vec{v}_i$
    \State $quality^0 \gets Q(\vec{v}_i)$
    \State $n \gets 1$
    \State $\vec{v}_i^n \gets \vec{v}_i^{\mathcal{L}}$ \Comment{Initialise vertex location using Laplacian smooth}
    \State $M_i^n \gets metric\_interpolation(\vec{v}_i^n)$
    \State $quality^n = Q(\vec{v}_i^n)$ \Comment{Calculate the new local quality for this relocation.}
    \While{$(n \leq max\_iteration) \And (quality_i^{n} - quality_i^0 < \sigma_q)$}
      \State $\vec{v}_i^{n+1} \gets (\vec{v}_i^n+\vec{v}_i^0)/2$
      \State $M_i^{n+1} \gets metric\_interpolation(\vec{v}_i^{n+1})$
      \State $quality^{n+1} \gets Q(\vec{v}_i^{n+1})$
      \State $n=n+1$
    \EndWhile
    \If{$quality_i^n - quality_i^0 > \sigma_q$} \Comment{Accept if local quality is improved}
      \State $\vec{v}_i \gets \vec{v}_i^n$
      \State $\mathbf{M_i} \gets \mathbf{M_i^n}$
    \EndIf
  \end{algorithmic}
\end{algorithm}

\section{Thread-level parallelism in mesh optimisation}
\label{sect:parallel}

To allow fine grained parallelisation of anisotropic mesh adaptation
we make extensive use of maximal independent sets. This approach was
first suggested in a parallel framework proposed by Freitag \etal
\cite{Freitag98thescalability}. However, while this approach avoids
updates being applied concurrently to the same neighbourhood, data
writes will still incur significant lock and synchronisation
overheads. For this reason we incorporate a deferred updates strategy,
described below, to minimise synchronisations and allow parallel
writes.

\subsection{Design choices}
\label{subsect:design}

Before presenting the adaptive algorithms, it is necessary to give
a brief description of the data structures used to store
mesh-related information. Following that, we present a set of
techniques which help us avoid hazards and data races and guarantee
fast and safe concurrent read/write access to mesh data.

\subsubsection{Mesh data structures}
\label{subsubsect:structures}
The minimal information necessary to represent a mesh is an
element-node list (we refer to it in this article as \verb=ENList=),
which is implemented in \PRAGMATIC as an STL vector of triplets of
vertex IDs (\verb=std::vector<int>=), and an array of vertex coordinates
(referred to as \verb=coords=), which is an STL vector of pairs of
coordinates (\verb=std::vector<double>=). Element \verb=eid= is comprised
of vertices \verb=ENList[3*eid]=, \verb=ENList[3*eid+1]= and \verb=ENList[3*eid+2]=,
whereas the $x$- and $y$-coordinates of vertex \verb=vid= are stored in
\verb=coords[2*vid]= and \verb=coords[2*vid+1]= respectively.

It is also necessary to store the metric tensor field. The field is
discretised node-wise and every metric tensor is a symmetric 2-by-2 matrix.
For each mesh node, we need to store three values for the tensor: two values
for the two on-diagonal elements and one value for the two off-diagonal elements.
Thus, \verb=metric= is an STL vector of triplets of metric tensor values
(\verb=std::vector<double>=). The three components of the metric at vertex
\verb=vid= are stored at \verb=metric[3*vid]=, \verb=metric[3*vid+1]= and
\verb=metric[3*vid+2]=.

All necessary structural information about the mesh can be extracted from
\verb=ENList=. However, it is convenient to create and maintain two
additional adjacency-related structures, the node-node adjacency list
(referred to as \verb=NNList=) and the node-element adjacency list
(referred to as \verb=NEList=). \verb=NNList= is implemented as an STL
vector of STL vectors of vertex IDs (\verb=std::vector< std::vector<int> >=).
\verb=NNList[vid]= contains the IDs of all vertices adjacent to vertex
\verb=vid=. Similarly, \verb=NEList= is implemented as an STL vector of
STL sets of element IDs (\verb=std::vector< std::set<int> >=) and
\verb=NEList[vid]= contains the IDs of all elements which vertex
\verb=vid= is part of.

It should be noted that, contrary to common approaches in other adaptive
frameworks, we do not use other adjacency-related structures such as
element-element or edge-edge lists. Manipulating these lists and maintaining
their consistency throughout the adaptation process is quite complex and
constitutes an additional overhead. Instead, we opted for the approach of
finding all necessary adjacency information on the fly using \verb=ENList=,
\verb=NNList= and \verb=NEList=.

\subsubsection{Colouring}
\label{subsubsect:colouring}
There are two types of hazards when running mesh optimisation
algorithms in parallel: structural hazards and data races. The term
{\em structural hazards} refers to the situation where an adaptive
operation results in invalid or non-conforming edges and elements. For
example, on the local patch in Figure \ref{fig:swapping_propagation},
if two threads flip edges $\overline{AD}$ and $\overline{BD}$ at the
same time, the result will be two new edges $\overline{AC}$ and
$\overline{BE}$ crossing each other. Structural hazards for all
adaptive algorithms are avoided by colouring a graph whose nodes are
defined by the mesh vertices and edges are defined by the mesh
edges. Maximal independent sets are readily selected by calculating
the intersection between the set of vertices of each colour and the
set of active vertices.

The fact that topological changes are made to the mesh means that
after an independent set has been processed the graph colouring has to
be recalculated. Therefore, a fast scalable graph colouring algorithm
is vital to the overall performance. In this work we use a parallel
colouring algorithm described by Gebremedhin \etal \cite{gebremedhin2000scalable}. This
algorithm can be described as having three stages: (a) initial
pseudo-colouring where vertices are coloured in parallel and invalid
colourings are possible; (b) loop over the graph to detect invalid
colours arising from the first stage; (c) the detected invalid colours
are resolved in serial. Between adaptive sweeps through independent
sets it is only necessary to execute stages (b) and (c) to resolve the
colour conflicts introduced by changes to the mesh topology.

\subsubsection{Deferred operations mechanism}
\label{subsubsect:deferred}
Data race conditions can appear when two or more threads try to update
the same adjacency list. An example can be seen in Figure
\ref{fig:edge_collapse_hazard}. Having coloured the mesh, two threads
are allowed to process vertices $V_B$ and $V_C$ at the same time
without structural hazards. However, \verb=NNList[= $V_A$ \verb=]= and
\verb=NEList[= $V_A$ \verb=]= must be updated. If both threads try to
update them at the same time there will be a data race which could
lead to data corruption. One solution could be a distance-2 colouring
of the mesh (a distance-$k$ colouring of $\mathcal{G}$ is a colouring
in which no two vertices share the same colour if these vertices are
up to $k$ edges away from each other or, in other words, if there is a
path of length $\leq k$ from one vertex to the other). Although this
solution guarantees a race-free execution, a distance-2 colouring
would increase the chromatic number, thereby reducing the size of the
independent sets and therefore the available parallelism. Therefore,
an alternative solution is sought.

In a shared-memory environment with \verb=nthreads= OpenMP threads,
every thread has a private collection of \verb=nthreads= lists, one
list for each OpenMP thread. When \verb=NNList[i]= or \verb=NEList[i]=
have to be updated, the thread does not commit the update immediately;
instead, it pushes the update back into the list corresponding to
thread with ID $tid=i\%nthreads$.  At the end of the adaptive
algorithm, every thread \verb=tid= visits the private collections of
all OpenMP threads (including its own), locates the list that was
reserved for \verb=tid= and commits the operations which are stored
there. This way, it is guaranteed that for any vertex $V_i$,
\verb=NNList[= $V_i$ \verb=]= and \verb=NEList[= $V_i$ \verb=]= will
be updated by only one thread. Because updates are not committed
immediately but are deferred until the end of the iteration of an
adaptive algorithm, we call this technique the {\em deferred
  updates}. A typical usage scenario is demonstrated in Algorithm
\ref{alg:deferred_updates}.

\begin{algorithm}[tbp]
\caption{Typical example of using the deferred updates mechanism}
\label{alg:deferred_updates}
\begin{verbatim}
typedef std::vector<Updates> DeferredOperationsList;
int nthreads = omp_get_max_threads();

// Create nthreads collections of deferred operations lists
std::vector< std::vector<DeferredOperationsList> > defOp;
defOp.resize(nthreads);

#pragma omp parallel
{
  // Every OMP thread executes
  int tid = omp_get_thread_num();
  defOp[tid].resize(nthreads);
  // By now, every OMP thread has allocated one list per thread
  
  // Execute one iteration of an adaptive algorithm in parallel
  // Defer any updates until the end of the iteration
  #pragma omp for
  for(int i=0; i<nVertices; ++i){
    execute kernel(i);
    // Update will be committed by thread i%nthreads
    // where the modulo avoids racing.
    defOp[tid][i%nthreads].push_back(some_update_operation);
  }
  
  // Traverse all lists which were allocated for thread tid
  // and commit any updates found
  for(int i=0; i<nthreads; ++i){
    commit_all_updates(defOp[i][tid]);
  }
}
\end{verbatim}
\end{algorithm}

\subsubsection{Worklists and atomic operations}
\label{subsubsect:worklists}
There are many cases where it is necessary to create a worklist of
items which need to be processed. An example of such a case is the
creation of the active sub-mesh in coarsening and swapping, as will be
described in Section \ref{subsubsect:coarsening_parallel}. Every
thread keeps a local list of vertices it has marked as active and all
local worklists have to be accumulated into a global worklist, which
essentially is the set of all vertices comprising the active sub-mesh.

One approach is to wait for every thread to exit the parallel loop and
then perform a prefix sum \cite{BlellochTR90} (also known as inclusive
scan or partial reduction in MPI terminology) on the number of vertices
in its private list. After that, every thread knows its index in the global
worklist at which it has to copy the private list. This method has the
disadvantage that every thread must wait for all other threads to exit the
parallel loop, synchronise with them to perform the prefix sum and finally
copy its private data into the global worklist. Profiling
data indicates that this way of manipulating worklists is a significant limiting factor
towards achieving good scalability.

Experimental evaluation showed that, at least on the Intel Xeon, a
better method is based on atomic operations on a global integer
variable which stores the size of the worklist needed so far. Every
thread which exits the parallel loop increments this integer
atomically while caching the old value. This way, the thread knows
immediately at which index it must copy its private data and
increments the integer by the size of this data, so that the next
thread which will access this integer knows in turn its index at which
its private data must be copied. Caching the old value before the
atomic increment is known in OpenMP terminology as {\em atomic
capture}. Support for atomic capture operations was introduced in
OpenMP 3.1. This functionality has also been supported by GNU
extensions (intrinsics) since GCC 4.1.2, known under the name
{\em fetch-and-add}. An example of using this technique is shown in
Algorithm \ref{alg:atomic_capture}.

Note the \verb=nowait= clause at the end of the \verb=#pragma omp for=
directive. A thread which exits the loop does not have to wait for the
other threads to exit. It can proceed directly to the atomic
operation. It has been observed that dynamic scheduling for OpenMP
for-loops is what works best for most of the adaptive loops in mesh
optimisation because of the irregular load balance across the
mesh. Depending on the nature of the loop and the chunk size, threads
will exit the loop at significantly different times. Instead of having
some threads waiting for others to finish the parallel loop, with this
approach they do not waste time and proceed to the atomic
increment. The profiling data suggests that the
overhead or spinlock associated with atomic-capture operations is insignificant.

\begin{algorithm}[tbp]
\caption{Example of creating a worklist using OpenMP's atomic capture operations.}
\label{alg:atomic_capture}
\begin{verbatim}
int worklistSize = 0; // Points to end of the global worklist 
std::vector<Item> globalWorklist;

// Pre-allocate enough space
globalWorklist.resize(some_appropriate_size);

#pragma omp parallel
{
  std::vector<Item> private_list;

  // Private list - no need to synchronise at end of loop.
  #pragma omp for nowait
  for(all items which need to be processed){
    do_some_work();
    private_list.push_back(item);
  }
  
  // Private variable - the index in the global worklist
  // at which the thread will copy the data in private_list.
  int idx;
  
  #pragma omp atomic capture
  {
    idx = worklistSize;
    worklistSize += private_list.size();
  }
  
  memcpy(&globalWorklist[idx], &private_list[0],
             private_list.size() * sizeof(Item));
  
}
\end{verbatim}
\end{algorithm}

\subsubsection{Reflection on alternatives}
\label{subsubsect:alternatives}

Our initial approach to dealing with structural hazards, data races
and propagation of adaptivity was based on a thread-partitioning
scheme in which the mesh was split into as many sub-meshes as there
were threads available. Each thread was then free to process
items inside its own partition without worrying about hazards and
races. Items on the halo of each thread-partition were locked (analogous to the MPI parallel strategy); those items
would be processed later by a single thread. However, this approach did not
result in good scalability for a number of reasons. Partitioning the mesh was a
significant serial overhead, which was incurred repeatedly as the
adaptive algorithms changed mesh topology and invalidated the
existing partitioning. In addition, the single-threaded phase of processing halo
items was another hotspot of this thread-partition approach. In line with Amdahl's law, these effects only become more pronounced as the number of threads is increased. For these reasons this thread-level domain decomposition approach was not pursued further.

\subsection{Refinement}
\label{subsubsect:refinement_parallel}

Every edge can be processed and refined without being affected by what
happens to adjacent edges. Being free from structural hazards, the only
issue we are concerned with is thread safety when updating mesh data
structures. Refining an edge involves the addition of a new vertex to the
mesh. This means that new coordinates and metric tensor values have to be
appended to \verb=coords= and \verb=metric= and adjacency information in
\verb=NNList= has to be updated. The subsequent element split leads to the
removal of parent elements from \verb=ENList= and the addition of new ones,
which, in turn, means that \verb=NEList= has to be updated as well.
Appending new coordinates to \verb=coords=, metric tensors to \verb=metric=
and elements to \verb=ENList= is done using the thread worklist strategy
described in Section \ref{subsubsect:worklists}, while updates to
\verb=NNList= and \verb=NEList= can be handled efficiently using the
deferred operations mechanism.

The two stages, namely edge refinement and element refinement, of our
threaded implementation are described in Algorithm
\ref{alg:edge_refinement_parallel} and Algorithm
\ref{alg:element_refinement_parallel}, respectively. The procedure
begins with the traversal of all mesh edges. Edges are accessed using
\verb=NNList=, \ie for each mesh vertex $V_i$ the algorithm visits $V_i$'s
neighbours. This means that edge refinement is a directed operation, as
edge $\overline{V_iV_j}$ is considered to be different from edge
$\overline{V_jV_i}$. Processing the same physical edge twice is
avoided by imposing the restriction that we only consider edges for
which $V_i$'s ID is less than $V_j$'s ID. If an edge is found to be
longer than desired, then it is split in the middle (in metric space)
and a new vertex $V_n$ is created. $V_n$ is associated with a pair of
coordinates and a metric tensor. It also needs an ID. At this stage,
$V_n$'s ID cannot be determined. Once an OpenMP thread exits the edge
refinement phase, it can proceed (without synchronisation with the other
threads) to fix vertex IDs and append the new data it created to the
mesh. The thread captures the number of mesh vertices $index=NNodes$
and increments it atomically by the number of new vertices it
created. After capturing the index, the thread can assign IDs to the
vertices it created and also copy the new coordinates and metric
tensors into \verb=coords= and \verb=metric=, respectively.

Before proceeding to element refinement, all split edges are
accumulated into a global worklist. For each split edge
$\overline{V_iV_j}$, the original vertices $V_i$ and $V_j$ have to be
connected to the newly created vertex $V_n$. Updating \verb=NNList=
for these vertices cannot be deferred. Most edges are shared between
two elements, so if the update was deferred until the corresponding
element were processed, we would run the risk of committing these
updates twice, once for each element sharing the edge. Updates can be
committed immediately, as there are no race conditions when accessing
\verb=NNList= at this point. Besides, for each split edge we find the
(usually two) elements sharing it. For each element, we record that
this edge has been split. Doing so makes element refinement much
easier, because as soon as we visit an element we will know
immediately how many and which of its edges have been split. An array
of length \verb=NElements= stores this type of information.

During mesh refinement, elements are visited in parallel and refined
independently. It should be noted that all updates to \verb=NNList=
and \verb=NEList= are deferred operations. After finishing the loop,
each thread uses the worklist method to append the new elements it
created to \verb=ENList=. Once again, no thread synchronisation is
needed.

This parallel refinement algorithm has the advantage of not requiring
any mesh colouring and having low synchronisation overhead as compared
with Freitag's task graph approach.

\begin{algorithm}[t]
  \caption{Edge-refinement.}
  \label{alg:edge_refinement_parallel}
  \begin{algorithmic}
   \State Global worklist of split edges $\mathcal{W}$, refined\_edges\_per\_element[NElements]
   \algblockdefx[BLOCK]{Start}{End}{\textbf{\#pragma omp parallel}}{}
   \Start
    \State $private:split\_cnt \gets 0, newCoords, newMetric, newVertices$
    \State \textbf{\#pragma omp for schedule(dynamic)}
    \ForAll{vertices $V_i$}
     \ForAll{vertices $V_j$ adjacent to $V_i$, $ID(V_i) < ID(V_j)$}
      \If{$length\ of\ edge\ \overline{V_iV_j} > L_{max}$}
       \State $V_n \gets$ new vertex of split edge $\overline{V_iV_nV_j}$; Append new
       \State coordinates, interpolated metric, split edge to $newCoords$,
       \State $newMetric$, $newVertices$; $split\_cnt \gets split\_cnt+1$
      \EndIf
     \EndFor
    \EndFor
    \algblockdefx[BLOCKS]{StartS}{EndS}{\textbf{\#pragma omp atomic capture}}{}
    \StartS
     \State $index \gets NNodes$; $NNodes \gets NNodes + splint\_cnt$
    \EndS
    Copy $newCoords$ into \verb=coords=, $newMetric$ into \verb=metric=
    \ForAll{edges $e_i \in newVertices$}
     \State $e_i=\overline{V_iV_nV_j}$; increment ID of $V_n$ by $index$
    \EndFor
    \State Copy $newVertices$ into $\mathcal{W}$
    \State \textbf{\#pragma omp barrier}
    \State \textbf{\#pragma omp parallel for schedule(dynamic)}
    \ForAll{Edges $e_i \in \mathcal{W}$}
     \State Replace $V_j$ with $V_n$ in \verb=NNList[=$V_i$\verb=]=; replace $V_i$ with $V_n$ in \verb=NNList[=$V_j$\verb=]=
     \State Add $V_i$ and $V_j$ to \verb=NNList[=$V_n$\verb=]=
     \ForAll{$elements\ E_i \in \{NEList[V_i] \cap NEList[V_j]\}$}
      \State Mark edge $e_i$ as refined in refined\_edges\_per\_element[$E_i$].
     \EndFor
    \EndFor
   \End
  \end{algorithmic}
\end{algorithm}

\begin{algorithm}[t]
  \caption{Element refinement phase}
  \label{alg:element_refinement_parallel}
  \begin{algorithmic}
    \algblockdefx[BLOCK]{Start}{End}{\textbf{\#pragma omp parallel}}{}
    \Start
     \State $private:newElements$
     \State \textbf{\#pragma omp for schedule(dynamic)}
     \ForAll{elements $E_i$}
      \State \Call{refine\_element}{$E_i$}
      \State Append additional elements to $newElements$.
     \EndFor
     \State Resize \verb=ENList=.
     \State Parallel copy of $newElements$ into \verb=ENList=.
    \End
  \end{algorithmic}
\end{algorithm}

\subsection{Coarsening}
\label{subsubsect:coarsening_parallel}

Because any decision on whether to collapse an edge is strongly
dependent upon what other edges are collapsing in the immediate
neighbourhood of elements, an operation task graph for coarsening has
to be constructed. Edge collapse is based on the removal of vertices,
\ie the elemental operation for edge collapse is the removal of a
vertex. Therefore, the operation task graph $\mathcal{G}$ is the mesh
itself.

Figure \ref{fig:edge_collapse_hazard} demonstrates what needs
to be taken into account in order to perform parallel coarsening
safely. It is clear that adjacent vertices cannot collapse
concurrently, so a distance-1 colouring of the mesh is sufficient in
order to avoid structural hazards. This colouring also enforces processing
of vertices topologically at least {\it every other one} which prevents
skewed elements forming during significant coarsening\cite{de1999parallel, li20053d}.

An additional consideration is that vertices which are two edges away
from each other share some common vertex $V_{common}$. Removing both
vertices at once means that $V_{common}$'s adjacency list will have to
be modified concurrently by two different threads, leading to data
races. These races can be avoided using the deferred operations
mechanism.

\fig{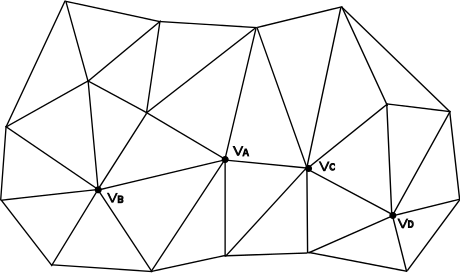}{Example of hazards when running edge
  collapse in parallel. $V_B$ is about to collapse onto $V_A$. The
  operation is executed by thread $T_1$. Clearly, $V_A$ cannot
  collapse at the same time. Additionally, $V_C$ cannot collapse
  either, because it affects $V_A$'s adjacency list. If a thread $T_2$
  executes the collapse operation collapse on $V_C$, then both $T_1$
  and $T_2$ will attempt to modify $V_A$'s adjacency list
  concurrently, which can lead to data corruption. This race can be
  eliminated using the deferred-updates mechanism.}{0.5}{fig:edge_collapse_hazard}

\begin{algorithm}[tbp]
  \caption{Edge collapse.}
  \label{alg:edge_collapse_parallel}
  \begin{algorithmic}
   \State Allocate $dynamic\_vertex, worklist$.
   \algblockdefx[BLOCK]{Start}{End}{\textbf{\#pragma omp parallel}}{}
   \Start
     \State \textbf{\#pragma omp for schedule(static)}
     \ForAll{vertices $V_i$}
       $dynamic\_vertex[V_i] \gets -2$
     \EndFor
     \State Colour mesh
     \Repeat
       \State \textbf{\#pragma omp for schedule(dynamic)}
       \ForAll{vertices $V_i$}
         \If{$dynamic\_vertex[V_i]==-2$}
           \State $dynamic\_vertex[V_i] \gets$\Call{coarsen\_identify}{$V_i$}
         \EndIf
       \EndFor
       \If{dynamic vertex count $==0$}
         \textbf{break}
       \EndIf
       \State $\mathcal{I}_m \gets$ maximal independent set of dynamic vertices
       \State \textbf{\#pragma omp for schedule(dynamic)}
       \ForAll{$V_i \in \mathcal{I}_m$}
         \State \Comment mark all neighbours for re-evaluation
         \ForAll{vertices $V_j \in$ NNList[$V_i$]}
           \State $dynamic\_vertex[V_j] \gets -2$
         \EndFor
         \State $dynamic\_vertex[V_i] \gets -1$
         \State \Call{coarsen\_kernel}{$V_i$}
       \EndFor
       \algblockdefx[BLOCKS]{StartS}{EndS}{\textbf{\#pragma omp single}}{}
       \State Commit deferred operations.
       \State Repair colouring
     \Until{\textbf{true}}
   \End
  \end{algorithmic}
\end{algorithm}

\begin{algorithm}[tbp]
  \caption{coarsen\_identify}
  \label{alg:coarsen_identify_kernel}
  \begin{algorithmic}
    \Procedure{coarsen\_identify}{$V_i$}
    \State $S_i \gets$ the set of all edges connected to $V_i$
    \State $S^0 \gets S_i$
    \Repeat
      \State $E_j \gets$ shortest edge in $S^j$
      \If{length of $E_j > L_{min}$} \Comment if shortest edge is of acceptable
        \State \Return -1 \Comment length, no edge can be removed
      \EndIf
      \State $V_t \gets$ the other vertex that bounds $E_j$
      \State evaluate collapse of $E_j$ with the collapse of $V_i$ onto $V_t$
      \If{($\forall$ edges $\in S_i \leq L_{max}$) \And ($\not \exists$ inverted elements)}
        \State \Return $V_t$
      \Else
        \State remove $E_j$ from $S^j$ \Comment $E_j$ is not a candidate for collapse
      \EndIf
    \Until{$S_i = \emptyset$}
    \EndProcedure
  \end{algorithmic}
\end{algorithm}

\begin{algorithm}[tbp]
  \caption{Coarsen\_kernel with deferred operations}
  \label{alg:coarsen_kernel}
  \begin{algorithmic}
    \Procedure{coarsen\_kernel}{$V_i$}
    \State $V_t \gets dynamic\_vertex[V_i]$
    \State $removed\_elements \gets$ \verb=NEList[=$V_i$\verb=]= $\cap$ \verb=NEList[=$V_t$\verb=]=
    \State $common\_patch \gets$ \verb=NNList[=$V_i$\verb=]= $\cap$ \verb=NNList[=$V_t$\verb=]=
    \ForAll{$E_i \in removed\_elements$}
    \State $V_o \gets $ the other vertex of $E_i=\widehat{V_iV_tV_o}$
    \State \verb=NEList[=$V_o$\verb=].erase(=$E_i$\verb=)= \Comment deferred operation
    \State \verb=NEList[=$V_t$\verb=].erase(=$E_i$\verb=)= \Comment deferred operation
    \State \verb=NEList[=$V_i$\verb=].erase(=$E_i$\verb=)=
    \State \verb=ENList[3*=$E_i$\verb=]= $\gets -1$ \Comment erase element by resetting its first vertex
    \EndFor
    
    \ForAll{$E_i \in$ NEList[$V_i$]}
    \State replace $V_i$ with $V_t$ in \verb=ENList[3*=$E_i$\verb=+{0,1,2}]=
    \State \verb=NEList[=$V_t$\verb=].add(=$E_i$\verb=)= \Comment deferred operation
    \EndFor
    \State remove $V_i$ from \verb=NNList[=$V_t$\verb=]= \Comment deferred operation
    \ForAll{$V_c \in common\_patch$}
    \State remove $V_i$ from \verb=NNList[=$V_c$\verb=]= \Comment deferred operation
    \EndFor
    \ForAll{$V_n \not \in common\_patch$}
    \State replace $V_i$ with $V_t$ in \verb=NNList[=$V_n$\verb=]=
    \State add $V_n$ to \verb=NNList[=$V_t$\verb=]= \Comment deferred operation
    \EndFor
    \State \verb=NNList[=$V_i$\verb=].clear()=
    \State \verb=NEList[=$V_i$\verb=].clear()=
    \EndProcedure
  \end{algorithmic}
\end{algorithm}

Algorithm \ref{alg:edge_collapse_parallel} illustrates a thread
parallel version of mesh edge collapse. Coarsening is divided into two
phases: the first sweep through the mesh identifies what edges are to
be removed, see Algorithm \ref{alg:coarsen_identify_kernel}; and
the second phase actually applies the coarsening operation, see
Algorithm \ref{alg:coarsen_kernel}. Function \textit{coarsen\_identify($V_i$)}
takes as argument the ID of a vertex $V_i$, decides whether any of the
adjacent edges can collapse and returns the ID of the target vertex
$V_t$ onto which $V_i$ should collapse (or a negative value if no
adjacent edge can be removed). \textit{coarsen\_kernel($V_i$)}
performs the actual collapse, \ie removes $V_i$ from the mesh, updates
vertex adjacency information and removes the two deleted elements
from the element list.

Parallel coarsening begins with the initialisation of array
$dynamic\_vertex$ which is defined as:
\begin{equation*}
dynamic\_vertex[V_i] =
\left\{
\begin{array}{ll}
-1 & \textrm{$V_i$ cannot collapsed},\\
-2 & \textrm{$V_i$ must be re-evaluated},\\
V_t & \textrm{$V_i$ is about to collapse onto $V_t$}.
\end{array}
\right.
\end{equation*}
At the beginning, the whole array is initialised to -2, so
that all mesh vertices will be considered for collapse.

In each iteration of the outer coarsening loop, $coarsen\_identify\_kernel$
is called for all vertices which have been marked for (re-)evaluation.
Every vertex for which $dynamic\_vertex[V_i] \geq 0$ is said to be
\textit{dynamic} or \textit{active}. At this point, a reduction in the
total number of active vertices is necessary to determine whether there is
anything left for coarsening or the algorithm should exit the loop.

Next up, we find the maximal independent set of active vertices
$\mathcal{I}_m$. Working with independent sets not only ensures safe
parallel execution, but also enforces the {\em every other vertex}
rule. For every active vertex $V_r \in \mathcal{I}_m$ which is about
to collapse, the local neighbourhood of all vertices $V_a$ formerly
adjacent to $V_r$ changed and target vertices $dynamic\_vertex[V_a]$
may not be suitable choices any more. Therefore, when $V_r$ is
erased, all its neighbours are marked for re-evaluation. This is how
propagation of coarsening is implemented.

Algorithm \ref{alg:coarsen_kernel} describes how the actual coarsening
takes place in terms of modifications to mesh data structures. Updates
which can lead to race conditions have been pointed out. These updates
are deferred until the end of processing of the independent set. Before
moving to the next iteration, all deferred operations are committed
and colouring is repaired because edge collapse may have introduced
inconsistencies.

\subsection{Swapping}
\label{subsubsect:swapping_parallel}

The data dependencies in edge swapping are virtually identical to those
of edge coarsening. Therefore, it is possible to reuse the same thread
parallel algorithm as for coarsening in the previous section with slight
modifications

In order to avoid maintaining edge-related data structures (\eg edge-node
list, edge-edge adjacency lists etc.), an edge can be expressed in terms
of a pair of vertices. Just like in refinement, we define an edge $E_{ij}$
as a pair of vertices $(V_i, V_j)$, with $ID(V_i) < ID(V_j)$. We say that
$E_{ij}$ is outbound from $V_i$ and inbound to $V_j$. Consequently, the edge
$E_{ij}$ can be marked for swapping by adding $V_j$ to $marked\_edges[V_i]$.
Obviously, a vertex $V_i$ can have more than one outbound edge, so
unlike $dynamic\_vertex$ in coarsening, $marked\_edges$ in swapping needs
to be a vector of sets (\verb=std::vector< std::set<int> >=).

The algorithm begins by marking all edges. It then enters a
loop which is terminated when no marked edges remain. The maximal
independent set $\mathcal{I}_m$ of active vertices is calculated. A
vertex is considered active if at least one of its outbound edges is
marked. Following that, threads process all active vertices of
$\mathcal{I}_m$ in parallel. The thread processing vertex $V_i$
visits all edges in $marked\_edges[V_i]$ one after the other and
examines whether they can be swapped, \ie whether the operation will
improve the quality of the two elements sharing that edge. It is easy
to see that swapping two edges in parallel which are outbound
from two independent vertices involves no structural hazards.

Propagation of swapping is similar to that of coarsening. Consider
the local patch in Figure \ref{fig:swapping} and assume that a thread
is processing vertex $V_0$. If edge $\overline{V_0V_1}$ is flipped,
the two elements sharing that edge change in shape and quality, so
all four edges surrounding those elements (forming the rhombus in
bold) have to be marked for processing. This is how propagation is
implemented in swapping.

One last difference between swapping and coarsening is that
$\mathcal{I}_m$ needs to be traversed more than once before
proceeding to the next one. In the same example as above, assume that
all edges adjacent to $V_0$ are outbound and marked. If edge
$\overline{V_0V_1}$ is flipped, adjacency information for $V_1$,
$V_2$ and $V_3$ has to be updated. These updates have to be deferred
because another thread might try to update the same lists at the same
time (\eg the thread processing edge $\overline{V_CV_1}$). However,
not committing the changes immediately means that the thread
processing $V_0$ has a stale view of the local patch. More precisely,
\verb=NEList[=$V_2$\verb=]= and \verb=NEList[=$V_3$\verb=]= are
invalid and cannot be used to find what elements edges
$\overline{V_0V_2}$ and $\overline{V_0V_3}$ are part of. Therefore,
these two edges cannot be processed until the deferred operations
have been committed. On the other hand, the rest of $V_0$'s outbound
edges are free to be processed. Once all threads have processed
whichever edges they can for all vertices of the independent set,
deferred operations are committed and threads traverse the
independent set again (up to two more times in 2D) to process what
had been skipped before.

\subsection{Smoothing}

Algorithm \ref{alg:parallel_loop} illustrates the colouring
based algorithm for mesh smoothing. In this algorithm the graph
$\mathcal{G}(\mathcal{V}, \mathcal{E})$ consists of sets of vertices
$\mathcal{V}$ and edges $\mathcal{E}$ that are defined by the vertices
and edges of the computational mesh. By computing a vertex colouring
of $\mathcal{G}$ we can define independent sets of vertices,
$\mathcal{V}^c$, where $c$ is a computed colour. Thus, all vertices in
$\mathcal{V}^c$, for any $c$, can be updated concurrently without any
race conditions on dependent data. This is clear from the definition
of the smoothing kernel in Section \ref{subsect:smoothing}. Hence, within a
node, thread-safety is ensured by assigning a different independent
set $\mathcal{V}^c$ to each thread.

\begin{algorithm}[t]
  \caption{Thread-parallel mesh smoothing}
  \label{alg:parallel_loop}
  \begin{algorithmic}
    \Repeat
    \State $relocate\_count \gets 0$
    \For{$colour = 1 \to k$}
      \State \textbf{\#pragma omp for schedule(static)}
      \ForAll{$i \in \mathcal{V}^c$}
       \State \Comment{$move\_success$ is {\tt true} if vertex was relocated,}
       \State $move\_success \gets smooth\_kernel(i)$ \Comment{{\tt false} otherwise.}
       \If{$move\_success$}
         \State $relocate\_count \gets relocate\_count + 1$
       \EndIf
      \EndFor
    \EndFor
    \Until{$(n \geq max\_iteration) \Or (relocate\_count = 0)$}
  \end{algorithmic}
\end{algorithm}

\section{Results}
\label{sect:benchmark}

In order to evaluate the parallel performance, an isotropic
mesh was generated on the unit square with using
approximately $200\times200$ vertices. A synthetic solution
$\psi$ is defined to vary in time and space:

\begin{equation}
  \psi(x, y, t) = 0.1\sin(50x+2\pi t/T) + \arctan(-0.1/(2x - \sin(5y+2\pi t/T))),
\end{equation}

where $T$ is the period. An example of the field at $t=0$ is
shown in Figure \ref{fig:field}. This is a good choice as a
benchmark as it contains multi-scale features and a shock
front. These are the typical solution characteristics where
anisotropic adaptive mesh methods excel.

\fig{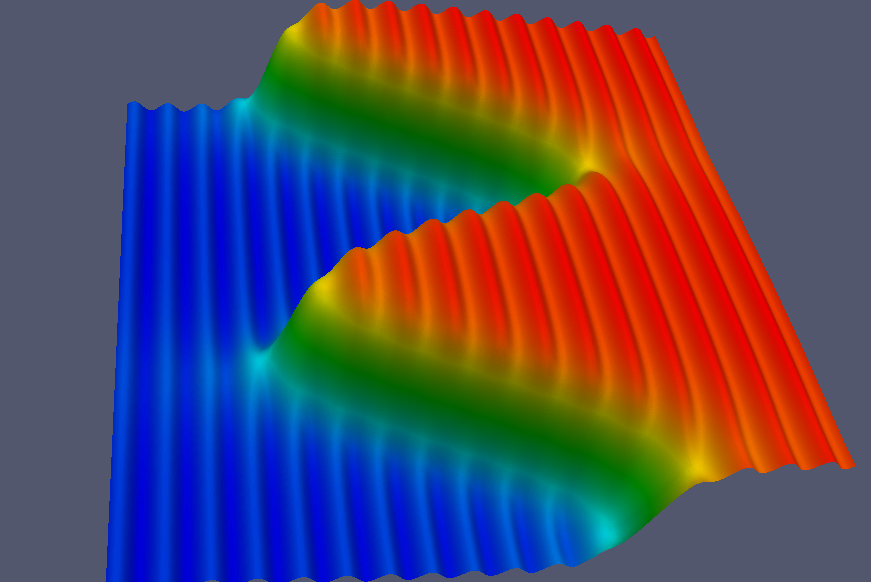}{Benchmark solution field.}{1.0}{fig:field}

Because mesh adaptation has a very irregular workload we simulate a
time varying scenario where $t$ varies from $0$ to $51$ in increments
of unity and we use the mean and standard deviations when reporting
performance results. To calculate the metric we used the
$L^{p=2}$-norm as described by \cite{chen2007optimal}. The number
of mesh vertices and elements maintains an average of approximately
$250k$ and $500k$ respectively. As the field evolves all of the adaptive
operations are heavily used, thereby giving an overall profile of the
execution time.

In order to demonstrate the correctness of the adaptive algorithm we
plot a histogram (Figure \ref{fig:quality_histogram}) showing the
quality of all element aggregated over all time steps. We can see that
the vast majority of the elements are of very quality. The lowest
quality element had a quality of $0.34$, and in total only $10$
elements out of $26$ million have a quality less than $0.4$.

\fig{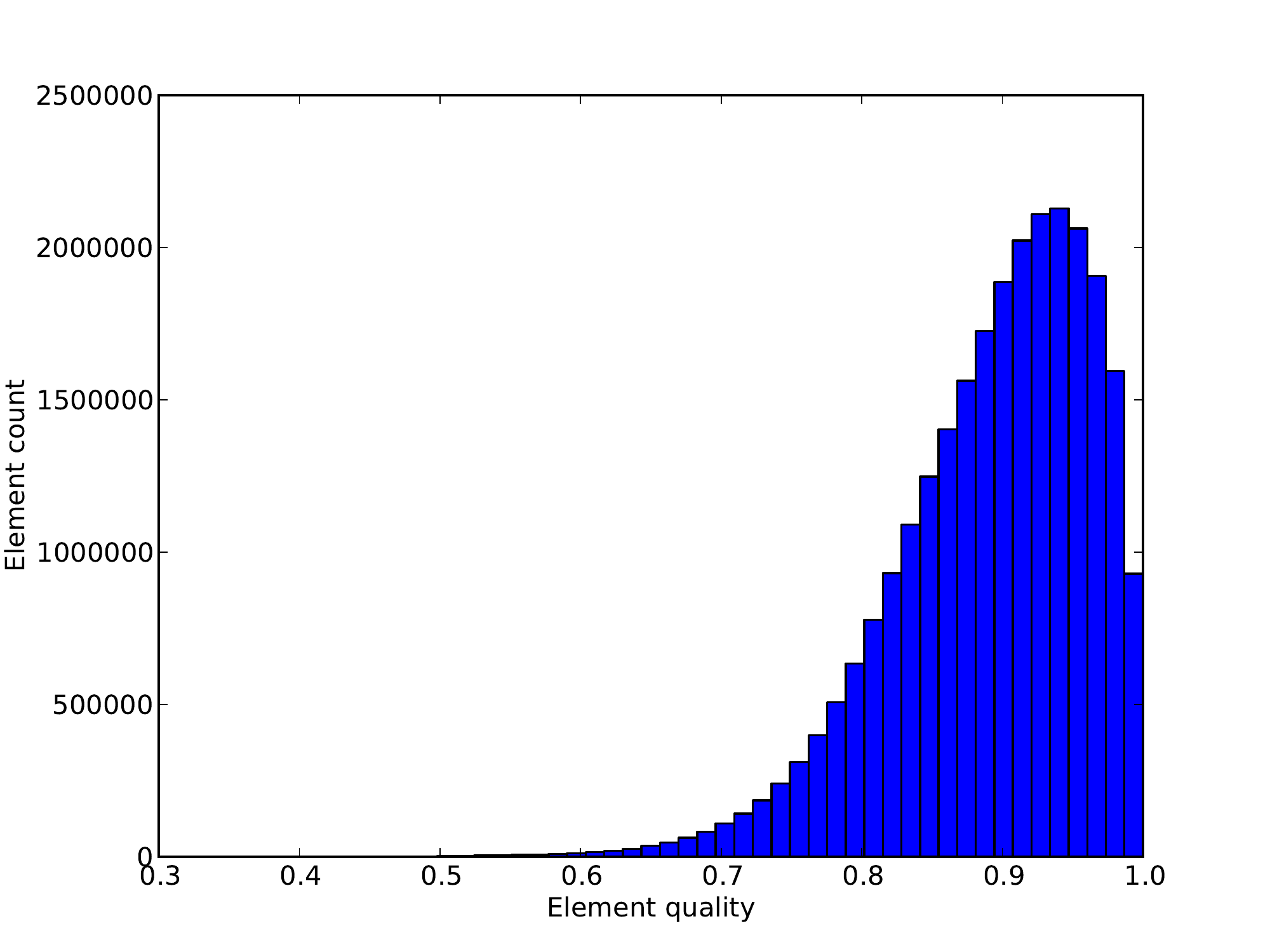}{Histogram of all element qualities aggregated over all time steps.}{1.0}{fig:quality_histogram}

The benchmarks were run on a Intel(R) Xeon(R) E5-2650 CPU.
The code was compiled using the Intel compiler suite,
version 13.0.0 and with the compiler flags {\tt -O3 -xAVX}.
In all cases thread-core affinity was used.

Figures \ref{fig:times}, \ref{fig:speedup} and
\ref{fig:efficiency} show the wall time, speedup and
efficiency of each phase of mesh adaptation. Simulations
using between 1 and 8 cores are run on a single socket while
the 16 core simulation runs across two CPU sockets and
thereby incurring NUMA overheads. From the results we can
see that all operations achieve good scaling, including
for the 16 core NUMA case. The dominant factors
limiting scaling are the number of synchronisations and
load-imbalances. Even in the case of mesh smoothing, which
involves the least data-writes, the relatively expensive
optimisation kernel is only executed for patches of elements
whose quality falls below a minimum quality tolerance. Indeed, the fact
that mesh refinement, coarsening and refinement are
comparable is very encouraging as it indicates that despite
the invasive nature of the operations on these relatively
complex data structures it is possible to get good
intra-node scaling.

\fig{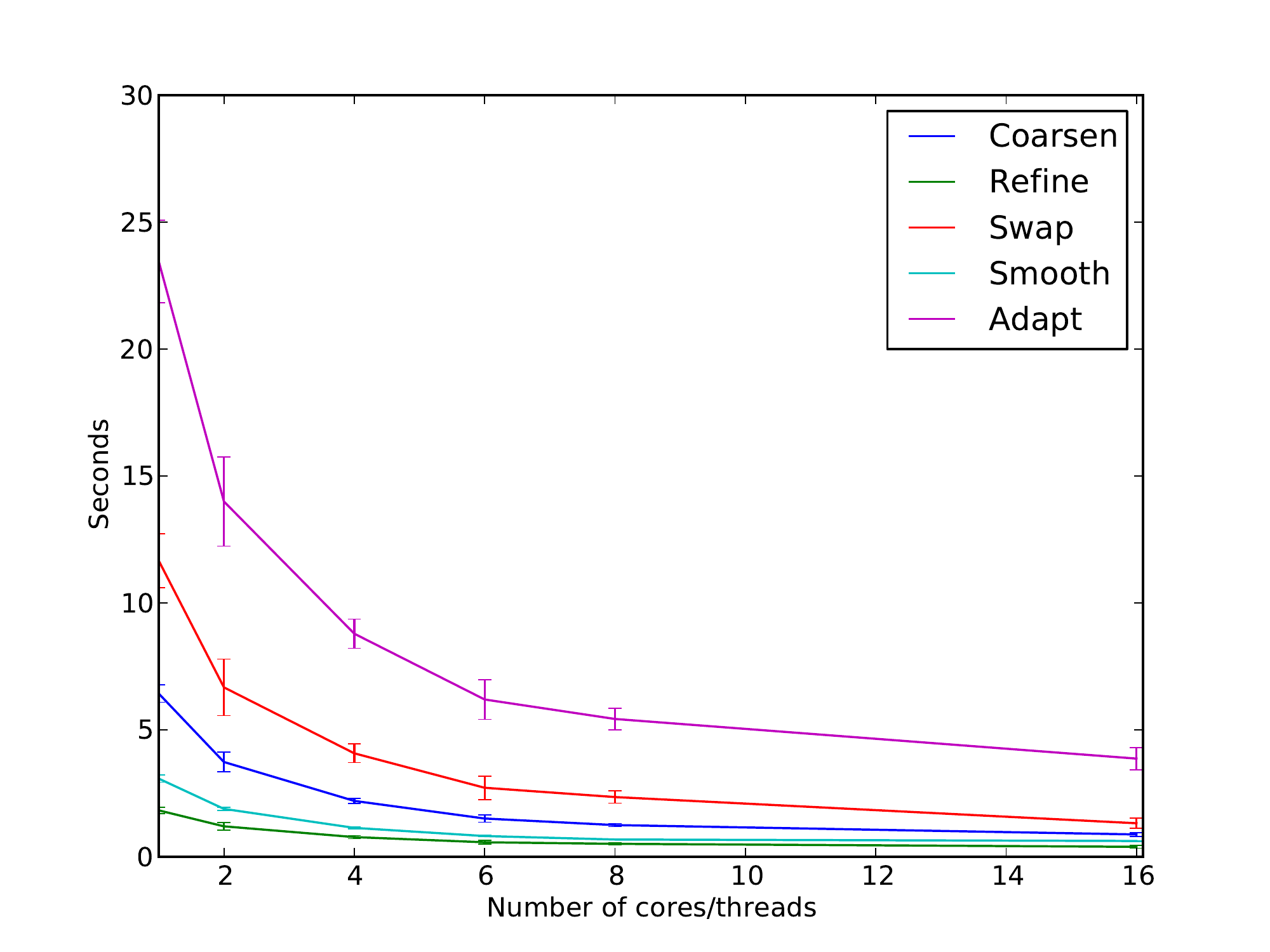}{Wall time for each phase of mesh adaptation.}{1.0}{fig:times}
\fig{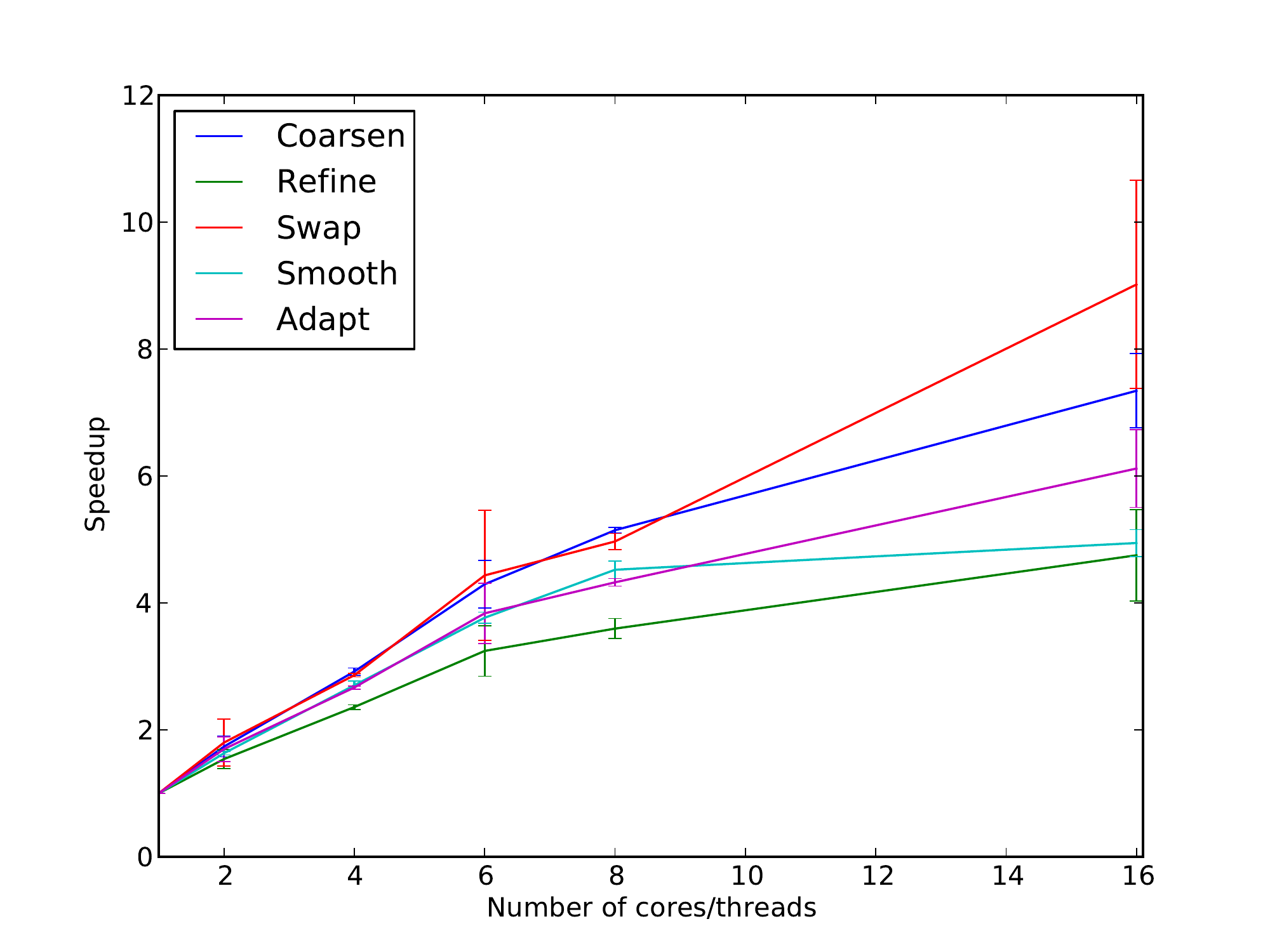}{Speedup profile for each phase of mesh adaptation.}{1.0}{fig:speedup}
\fig{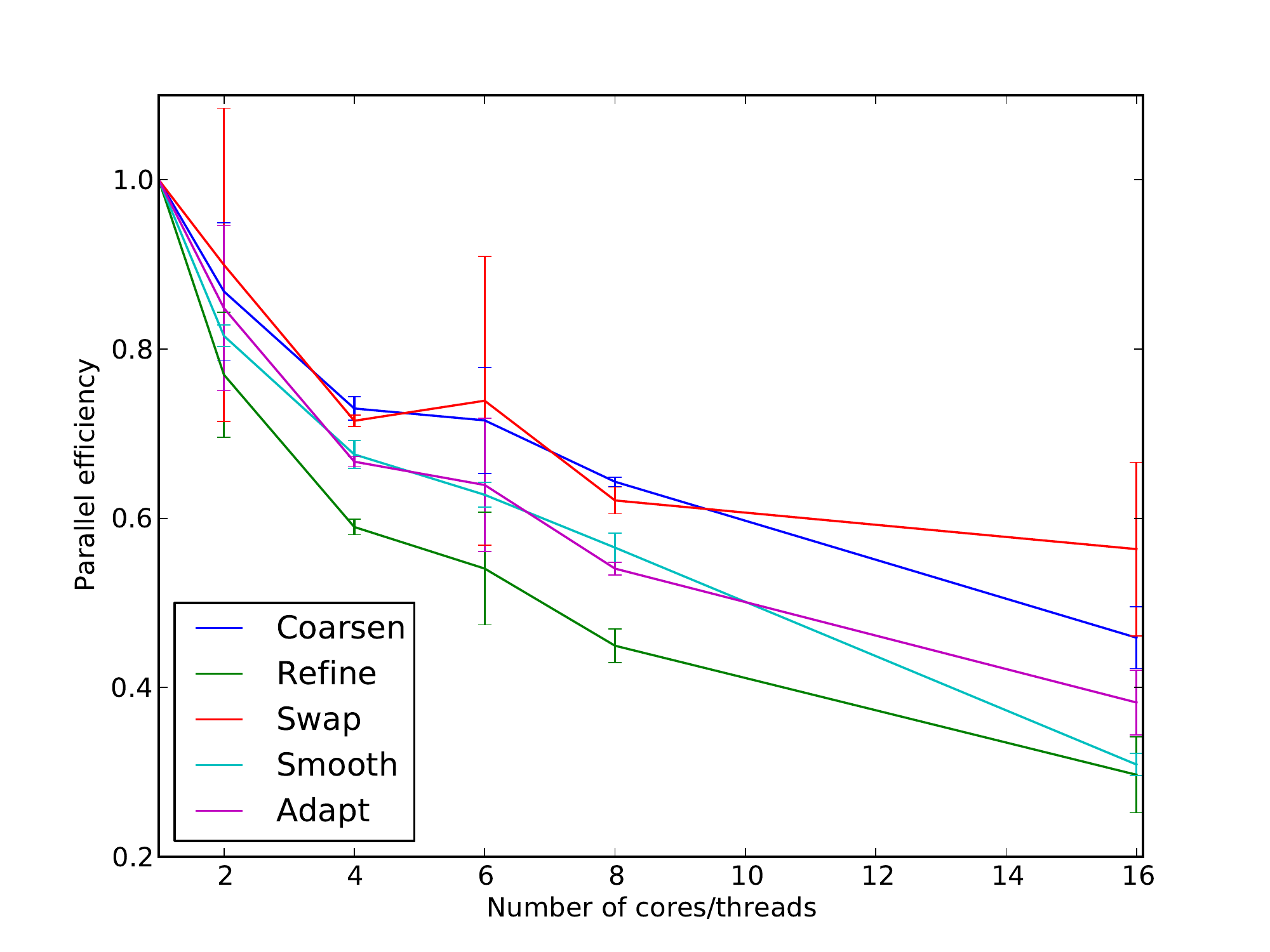}{Parallel efficiency profile for each phase of mesh adaptation.}{1.0}{fig:efficiency}

\section{Conclusion}

This paper is the first to examine the scalability of anisotropic mesh
adaptivity using a thread-parallel programming model and to explore new parallel
algorithmic approaches to support this model. Despite the complex data
dependencies and inherent load imbalances we have shown it is possible to
achieve practical levels of scaling. To achieve this two key ingredients were
required. The first was to use colouring to identify maximal independant sets of
tasks that would be performed concurrently. In principle this facilitates
scaling up to the point that the number of elements of the independant set is
equal to the number of available threads. The second important factor
contributing to the scalability was the use of worklists and deferred whereby
updates to the mesh are added to worklists and applied in parallel at a later
phase of an adaptive sweep. This avoids the majority of serial overheads
otherwise incurred with updating mesh data structures. 

While the algorithms presented are for 2D anisotropic mesh adaptivity, we
believe many of the algorithmic details carry over to the 3D case as the 
challenges associated with exposing a sufficient degree of parallelism are very
similar.

\section{Acknowledgments}
The authors would like to thank Fujitsu Laboratories of Europe Ltd. and
EPSRC grant no. EP/I00677X/1 for supporting this work.

\bibliographystyle{elsarticle-num}
\bibliography{references}

\end{document}